\documentclass[aip,jcp,reprint,superscriptaddress,nofootinbib]{revtex4-2}

\usepackage{amsmath,amssymb,bm}
\usepackage{xcolor}
\usepackage[normalem]{ulem}
\usepackage{graphicx}
\usepackage{hyperref}

\usepackage{lineno}

\newif\iferbedit
\erbeditfalse

\iferbedit
  
  \newcommand{\erbdel}[1]{\textcolor{red}{\sout{#1}}}
  
\else
  
  \newcommand{\erbdel}[1]{}
  
\fi

\newcommand{\Tr}{\operatorname{Tr}}

\begin{document}

\title{Geometric Response in Open Quantum Systems: Coherence, Curvature, and Susceptibility}

\author{Eric R. Bittner}
\email{ebittner@central.uh.edu}
\affiliation{Department of Physics, University of Houston, Houston, Texas 77204, USA}
\affiliation{Institut Courtois \& D\'epartement de physique, Universit\'e de Montr\'eal, 1375 Avenue Th\'er\`ese-Lavoie-Roux, Montr\'eal H2V~0B3, Qu\'ebec, Canada}

\author{Carlos~Silva-Acu\~na}
\email{carlos.silva@umontreal.ca}
\affiliation{School of Chemistry and Biochemistry, Georgia Institute of Technology, 901 Atlantic Drive, Atlanta, GA~30332, United~States}
\affiliation{Institut Courtois \& D\'epartement de physique, Universit\'e de Montr\'eal, 1375 Avenue Th\'er\`ese-Lavoie-Roux, Montréal, Qu\'ebec H2V~0B3, Canada}
\affiliation{Departamento de F\'isica Aplicada, Centro de Investigaci\'on y de Estudios Avanzados del Instituto Polit\'ecnico Nacional (CINVESTAV), 97310 M\'erida, Yucat\'an, M\'exico}

\date{\today}
\begin{abstract}
We develop a response-geometric framework for open quantum 
systems by decomposing the stationary-state response tensor 
into symmetric and antisymmetric sectors. The symmetric 
sector defines a metric-like response tensor governing local 
susceptibility, while the antisymmetric sector defines a 
curvature two-form associated with nonreciprocal response 
and geometric work. Using linear response theory and the 
quantum regression theorem, we show that the work curvature 
is precisely the antisymmetric component of the response 
tensor, establishing a fluctuation--response relation that 
extends the geometric structure of equilibrium thermodynamics 
to nonequilibrium steady states. This decomposition reveals 
a response geometry containing both metric and symplectic 
sectors. In equilibrium, the antisymmetric sector vanishes 
as a consequence of reciprocity, yielding the familiar 
purely metric geometry of thermodynamic response. Open 
quantum systems admit a broader structure in which 
reciprocal and nonreciprocal response coexist on the same 
control manifold. To illustrate these ideas, we analyze a 
driven dissipative qubit subject to pure dephasing. The 
model demonstrates that finite curvature does not require 
strong driving, non-Markovian dynamics, or engineered 
reservoirs. Instead, the antisymmetric response emerges 
from the incompatibility between the Hamiltonian eigenbasis 
and the pointer basis selected by the environment. 
Comparison with the Bures information metric reveals that 
response geometry and information geometry characterize 
distinct physical properties of the same stationary-state 
manifold: the former governs susceptibility and geometric 
work, whereas the latter quantifies state distinguishability. 
These results identify geometric work as a measurable 
manifestation of nonreciprocal response and establish 
response geometry as a natural framework for nonequilibrium 
open quantum systems.
\end{abstract}

\maketitle

\section{Introduction}
\label{sec:intro}

Geometry has long played a central role in our understanding of
thermodynamic and quantum systems. In equilibrium thermodynamics,
geometric structures arise naturally from the Hessian of
thermodynamic potentials, leading to the Weinhold and Ruppeiner
metrics and providing a geometric description of fluctuations,
susceptibilities, stability, and critical phenomena
\cite{Weinhold1975,Ruppeiner1995}. In quantum mechanics,
geometric concepts such as Berry phases, quantum geometric tensors,
and quantum Fisher information characterize the response of quantum
states to parameter variation and provide fundamental measures of
distinguishability and sensitivity
\cite{Berry1984,ProvostVallee1980,Zanardi2007,Gu2010}.
Together, these developments suggest that response itself admits a
natural geometric interpretation.

The distinction between equilibrium and nonequilibrium response may
be understood geometrically. In equilibrium thermodynamics,
response is completely characterized by a symmetric metric
structure. Because equilibrium response derives from a state
function, reciprocity relations guarantee that the associated
antisymmetric sector vanishes identically and no geometric work
can accumulate over a closed cycle. In this sense, equilibrium
thermodynamics is fundamentally metric.

Driven open quantum systems depart from this paradigm.
Stationary states generally do not derive from an equilibrium
potential, and the response tensor need not be integrable.
Consequently, an antisymmetric response sector may emerge
alongside the familiar symmetric response sector. The resulting
response geometry contains both a metric component associated
with local susceptibility and a curvature component associated
with directed transport, response circulation, and quasistatic
work.

Recent years have witnessed growing interest in geometric
approaches to nonequilibrium quantum systems. Geometric phases and
quantum geometric tensors have been used extensively to
characterize quantum criticality, adiabatic transport, and
topological matter
\cite{Berry1984,ProvostVallee1980,Zanardi2007,Gu2010}.
Information-geometric methods based on the Bures and quantum Fisher
metrics have emerged as powerful tools for analyzing quantum phase
transitions, quantum metrology, and parameter estimation
\cite{Bures1969,Uhlmann1976,BraunsteinCaves1994,
ProvostVallee1980,Zanardi2007,Paris2009}. More recently,
information geometry has been extended to nonequilibrium steady
states, where geometric measures derived from stationary density
matrices have been shown to characterize transitions and critical
behavior in driven dissipative systems
\cite{Lacerda:2025aa,Bettmann:2025aa}. At the same time,
thermodynamic geometry has continued to provide important insights
into fluctuations, critical phenomena, and response in both
classical and quantum systems
\cite{Weinhold1975,Ruppeiner1995,Brandner2020,
Scandi2019,Abiuso2020,TerrenAlonso2022,Bhandari2020}.

Geometric structures have also begun to appear in the study of
open quantum systems and nonequilibrium steady states. Geometric
phases for mixed states, dissipative quantum transport, and
nonequilibrium geometric pumping have revealed that dissipation
can generate geometric phenomena absent in isolated systems
\cite{Buric2009,Carollo2018,Avron2000,Ren2010}. In parallel, our
recent work demonstrated that quasistatic work in open quantum
steady states is governed by a curvature two-form defined on the
control manifold, establishing a geometric theory of work in
driven dissipative systems \cite{Bittner:2026aa}.
The resulting work is determined by the flux of a curvature
through the area enclosed by a cycle, providing an open-system
analogue of geometric transport in thermodynamics.

The existence of a work curvature naturally raises a broader
question: what is the response-theoretic origin of this
curvature? In equilibrium statistical mechanics, fluctuations
and response are connected through fluctuation–dissipation
relations. Symmetric correlation functions govern variances and
susceptibilities, while antisymmetric correlations encode
dynamical response and transport. This observation suggests
that the curvature associated with geometric work may represent
the antisymmetric sector of a more general stationary-state
response geometry.

In this work we show that the work curvature previously
identified in nonequilibrium steady states is precisely the
antisymmetric sector of the stationary-state response tensor.
By decomposing the response tensor into symmetric and
antisymmetric components, we obtain a natural separation
between reciprocal and nonreciprocal response. The symmetric
sector defines a metric-like response tensor governing local
susceptibility, while the antisymmetric sector defines a
curvature two-form governing geometric work and response
circulation. Using linear response theory and the quantum
regression theorem, we derive an explicit fluctuation–response
relation connecting the curvature to dynamical correlation
functions.

This decomposition reveals a geometric structure that extends
classical thermodynamic geometry. Equilibrium systems possess
only the symmetric sector because reciprocity eliminates
antisymmetric response. Open quantum systems admit a broader
response geometry containing both metric and curvature
contributions. In this sense, nonequilibrium steady states
naturally acquire both Riemannian and symplectic
characteristics.

To illustrate these ideas, we analyze a driven dissipative
qubit subject to pure dephasing. This model demonstrates that
antisymmetric response does not require strong driving,
exceptional points, engineered reservoirs, or non-Markovian
dynamics. Instead, the curvature emerges from the
incompatibility between the Hamiltonian eigenbasis and the
pointer basis selected by the environment. We compare the
resulting response geometry with the Bures information
geometry of the same steady-state manifold and show that the
two structures encode distinct physical information: response
geometry characterizes susceptibility, nonreciprocity, and
geometric work, whereas information geometry characterizes
state distinguishability and statistical sensitivity.

The coexistence of metric and curvature sectors suggests a
deeper geometric organization. In particular, the response
decomposition naturally produces both metric and symplectic
structures on the control manifold, inviting comparison with
K\"ahler geometry, where compatible metric and symplectic
forms coexist on a common manifold. While the present work
does not construct an explicit complex structure, it
identifies the geometric ingredients from which such a
framework might emerge.

That nonequilibrium steady states carry this two-sector
geometric structure is not an isolated observation.
Beyen, Khodabandehlou, and Maes~\cite{Beyen:2026} have
shown contemporaneously and independently that quasistatic
response in nonequilibrium steady states is governed by a
Berry curvature whose nonvanishing signals the breakdown of
thermodynamic Maxwell relations. Their construction,
formulated for classical Markov jump processes, derives the
curvature from Liouvillian eigenvectors via a counting-field
formalism. The present work arrives at the same geometric
structure from a different direction. Rather than working
through an eigendecomposition of the Liouvillian, we
identify the work curvature directly as the antisymmetric
sector of the stationary-state response tensor --- a
construction that requires no spectral information and
extends naturally to open quantum systems. The physical
origin of the curvature in our setting is basis
incompatibility between the Hamiltonian eigenbasis and the
environment-selected pointer basis, a mechanism with no
classical analogue. The convergence of these two approaches
reinforces the conclusion that nonreciprocal response and
geometric work are robust features of nonequilibrium steady
states, classical and quantum alike.

The remainder of this paper is organized as follows.
Section~\ref{sec:response} develops the response-geometric
framework and derives the curvature from linear response
theory and the quantum regression theorem.
Section~\ref{sec:model} introduces the driven dissipative
qubit and compares the resulting response geometry with the
Bures information geometry of the stationary-state manifold.
Section~\ref{sec:discussion} concludes with a 
discussion of  the broader
implications for fluctuation–response theory,
nonequilibrium thermodynamics, and experimentally accessible
geometric observables. 
\section{Response Geometry and the Extension of
Fluctuation--Dissipation Theory}
\label{sec:response}
The fluctuation--dissipation theorem (FDT) provides one of the
central connections of statistical physics, relating the
response of a system to its fluctuations through equilibrium
correlation functions \cite{Kubo1957,Kubo1966}. In equilibrium
thermodynamics, response coefficients derive from a scalar
thermodynamic potential and therefore satisfy reciprocity
relations. Consequently, the geometry of equilibrium
thermodynamics is fundamentally metric in character, as
embodied by the Weinhold and Ruppeiner constructions
\cite{Weinhold1975,Ruppeiner1995}.
Open quantum systems extend this picture. Stationary states
need not be Gibbs states, response coefficients need not obey
reciprocity, and the response tensor may contain an
independent antisymmetric sector. The central result of this
section is that this antisymmetric response tensor is
precisely the work curvature governing geometric work in
nonequilibrium steady states
\cite{Bittner:2026aa}. This identification
provides a geometric extension of the fluctuation--dissipation
theorem and reveals that coherent open quantum systems possess
a Poisson sector of response geometry absent in equilibrium
thermodynamics. Classical thermodynamic geometry emerges as
the integrable limit in which this antisymmetric sector
vanishes identically.
\subsection{Response Tensor and Equilibrium Geometry}
Consider a family of stationary density matrices
\begin{equation}
\rho_{\rm ss}(\boldsymbol{\lambda}),
\end{equation}
parameterized by externally controlled variables
\begin{equation}
\boldsymbol{\lambda}
=
(\lambda^1,\lambda^2,\ldots).
\end{equation}
Examples include magnetic fields, detunings, interaction
strengths, driving amplitudes, cavity frequencies, or other
externally controlled Hamiltonian parameters. Together these
coordinates define a control manifold on which the stationary
state evolves.
Associated with each control parameter is a generalized force
\begin{equation}
O_\mu
=
\frac{\partial H}
{\partial\lambda^\mu},
\label{eq:generalized_force}
\end{equation}
which generalizes familiar thermodynamic conjugate variables
such as pressure,
$P=-\partial H/\partial V$,
or magnetization,
$M=-\partial H/\partial B$.
The stationary response tensor is defined by
\begin{equation}
\chi_{\mu\nu}
=
\frac{\partial}
{\partial\lambda^\nu}
\langle O_\mu\rangle_{\rm ss},
\label{eq:chi_def}
\end{equation}
where
\begin{equation}
\langle O_\mu\rangle_{\rm ss}
=
{\rm Tr}
\left[
\rho_{\rm ss} O_\mu
\right].
\end{equation}
The response tensor admits the decomposition
\begin{equation}
\chi_{\mu\nu}
=
G_{\mu\nu}
+
\chi_{\mu\nu}^{(-)},
\end{equation}
where
\begin{equation}
G_{\mu\nu}
=
\frac12
\left(
\chi_{\mu\nu}
+
\chi_{\nu\mu}
\right)
\label{eq:Gdef}
\end{equation}
and
\begin{equation}
\chi_{\mu\nu}^{(-)}
=
\frac12
\left(
\chi_{\mu\nu}
-
\chi_{\nu\mu}
\right).
\end{equation}
The symmetric tensor $G_{\mu\nu}$ describes the local
susceptibility structure of the stationary state and reduces
to the familiar equilibrium response tensor in the Gibbs
limit.
For an equilibrium Gibbs state,
\begin{equation}
\rho_{\rm eq}
=
\frac{e^{-\beta H}}
{Z},
\end{equation}
the generalized forces derive from a scalar free-energy
potential,
\begin{equation}
\langle O_\mu\rangle
=
\partial_\mu F_{\rm th},
\end{equation}
and therefore satisfy the reciprocity relation
\begin{equation}
\chi_{\mu\nu}
=
\partial_\nu\partial_\mu F_{\rm th}
=
\partial_\mu\partial_\nu F_{\rm th}
=
\chi_{\nu\mu}.
\end{equation}
Consequently,
\begin{equation}
\chi_{\mu\nu}^{(-)}=0.
\end{equation}
The fluctuation--dissipation theorem therefore probes only
the symmetric response tensor $G_{\mu\nu}$, which forms the
geometric foundation of equilibrium thermodynamics.

\subsection{Work Curvature as Antisymmetric Response}
In recent work
\cite{Bittner:2026aa}
we showed that quasistatic work in open quantum steady states
is governed by a geometric connection defined on the control
manifold,
\begin{equation}
A
=
A_\mu d\lambda^\mu,
\end{equation}
with components
\begin{equation}
A_\mu
=
{\rm Tr}
\left[
\rho_{\rm ss}
\frac{\partial H}
{\partial\lambda^\mu}
\right]
=
\langle O_\mu\rangle_{\rm ss}.
\label{eq:A_mu}
\end{equation}
The work performed along a quasistatic path
$\mathcal C$ is
\begin{equation}
W
=
\int_{\mathcal C}
A_\mu
d\lambda^\mu.
\end{equation}
The corresponding curvature is
\begin{equation}
F_{\mu\nu}
=
\partial_\mu A_\nu
-
\partial_\nu A_\mu.
\label{eq:F_def}
\end{equation}
Using Eq.~(\ref{eq:chi_def}) immediately gives
\begin{equation}
F_{\mu\nu}
=
\chi_{\nu\mu}
-
\chi_{\mu\nu}
=
-2\chi_{\mu\nu}^{(-)}.
\label{eq:F_chi_relation}
\end{equation}
Equation~(\ref{eq:F_chi_relation}) is the central result of 
this paper. Geometric work and nonreciprocal response are not 
analogous phenomena --- they are the same object. The curvature 
governing work accumulation over a closed cycle is precisely the 
antisymmetric sector of the stationary-state response tensor, 
requiring no additional geometric construction beyond response 
theory itself.

Physically, a nonzero curvature signifies that stationary-state 
response is nonreciprocal. Perturbing $\lambda^\nu$ and measuring 
the induced change in $O_\mu$ does not, in general, produce the 
same result as perturbing $\lambda^\mu$ and measuring $O_\nu$. 
The curvature $F_{\mu\nu}$ serves both as the local density of 
geometric work and as a measurable signature of non-integrable 
response. The symmetric tensor $G_{\mu\nu}$ governs reciprocal 
susceptibility and sets the metric of the steady-state manifold 
--- the local cost of quasistatic driving.

In equilibrium, response derives from a scalar free-energy 
potential and reciprocity holds exactly. The antisymmetric sector 
vanishes, $F_{\mu\nu} = 0$, and the geometry is purely metric 
--- the Weinhold and Ruppeiner constructions are recovered as 
special cases. Open quantum systems break this restriction. The 
stationary state need not derive from any potential, reciprocity 
need not hold, and both geometric sectors coexist on the same 
control manifold. Crucially, $F_{\mu\nu}$ vanishes whenever both 
$\lambda^\mu$ and $\lambda^\nu$ are parameters of the system 
Hamiltonian $H$; the antisymmetric sector activates only when 
the control manifold extends into bath parameter space. The 
geometry is therefore partitioned: Hamiltonian parameter 
directions span a purely Riemannian submanifold, while bath 
parameter directions introduce a symplectic sector. Geometric 
work is the physical signature of that coexistence.

\subsection{Kubo and Quantum Regression Representation}

The relation in Eq.~(\ref{eq:F_chi_relation}) also has a
microscopic representation in terms of causal response
functions. Let
\begin{equation}
\chi_{\mu\nu}^{R}(t)
=
-\frac{i}{\hbar}
\Theta(t)
\left\langle
[O_\mu(t),O_\nu(0)]
\right\rangle_{\rm ss}
\end{equation}
denote the retarded stationary response function, where the
time evolution is generated by the Liouvillian of the open
system and the average is taken in the stationary state. The
zero-frequency response is
\begin{equation}
\chi_{\mu\nu}^{R}(0)
=
\int_0^\infty dt\,
\chi_{\mu\nu}^{R}(t).
\end{equation}

For Hamiltonian controls with a fixed dissipator, the
quantum regression theorem gives the stationary response
tensor in the form
\begin{equation}
\chi_{\mu\nu}
=
\chi_{\mu\nu}^{R}(0)
+
\left\langle
\partial_\nu O_\mu
\right\rangle_{\rm ss}.
\label{eq:chi_kubo}
\end{equation}
The second term is symmetric under
$\mu\leftrightarrow\nu$ for a smooth Hamiltonian, since
\begin{equation}
\partial_\nu O_\mu
=
\partial_\nu\partial_\mu H
=
\partial_\mu\partial_\nu H
=
\partial_\mu O_\nu,
\end{equation}
it cancels from the antisymmetric response. Hence,
\begin{equation}
F_{\mu\nu}
=\chi_{\nu\mu}^{R}(0)-\chi_{\mu\nu}^{R}(0).
\label{eq:F_retarded}
\end{equation}
Equivalently,
\begin{equation}
F_{\mu\nu}
=
-\frac{i}{\hbar}
\int_0^\infty dt\,
\left[
\left\langle
[O_\nu(t),O_\mu(0)]
\right\rangle_{\rm ss}
-
\left\langle
[O_\mu(t),O_\nu(0)]
\right\rangle_{\rm ss}
\right].
\label{eq:F_QRT_precise}
\end{equation}

Equation~(\ref{eq:F_retarded}) is the response-theoretic
form of the work curvature: the curvature is the
antisymmetric zero-frequency retarded response. 
As a result, the
same object that generates geometric work is also the
nonreciprocal sector of the Kubo response. The symmetric
sector $G_{\mu\nu}$ is controlled by the reciprocal part of
the response and, in equilibrium, is related to fluctuations
by the usual fluctuation--dissipation theorem.

\subsection{Poisson Geometry of Response}
Since the work curvature is the exterior derivative of the
work connection,
\begin{equation}
F=dA,
\end{equation}
it defines a closed two-form on the control manifold,
\begin{equation}
F
=
\frac12
F_{\mu\nu}
d\lambda^\mu\wedge d\lambda^\nu.
\end{equation}
The appearance of a closed antisymmetric two-form is
significant because it introduces a geometric structure that
has no analogue in equilibrium thermodynamics. Whereas the
symmetric response tensor $G_{\mu\nu}$ describes local
susceptibility, the curvature $F_{\mu\nu}$ governs the
circulation of response around closed paths in parameter
space. In this sense, $G_{\mu\nu}$ characterizes local
response, while $F_{\mu\nu}$ characterizes response
transport.

Whenever the curvature is nondegenerate on a connected
region of parameter space, the corresponding two-form
defines a local symplectic structure. Its inverse, then, defines a Poisson tensor
\begin{equation}
\Pi^{\mu\nu}
=
(F^{-1})^{\mu\nu},
\end{equation}
and an associated bracket
\begin{equation}
\{f,h\}_F
=
\Pi^{\mu\nu}
\partial_\mu f
\partial_\nu h.
\end{equation}
Thus the antisymmetric response tensor does more than
quantify geometric work: it endows the control manifold with
a Poisson geometry.
This structure is absent in equilibrium thermodynamics.
Because Gibbs states derive from a scalar thermodynamic
potential, reciprocity implies
$F_{\mu\nu}=0$,
and the geometry is purely metric. Open quantum systems
extend this picture by introducing a Poisson sector
associated with nonreciprocal response.

A particularly important limit occurs when the stationary
state is diagonal in the Hamiltonian eigenbasis,
\begin{equation}
[\rho_{\rm ss},H]=0.
\end{equation}
In this case the stationary state is aligned with the
energetic structure of the system, the response becomes
integrable, and the antisymmetric sector vanishes. The
resulting geometry is purely metric.

Open quantum systems admit a more general situation in which
the preferred basis selected by dissipation does not coincide
with the Hamiltonian eigenbasis. This pointer--Hamiltonian
misalignment generates stationary-state coherence,
\begin{equation}
[\rho_{\rm ss},H]\neq0,
\end{equation}
and activates the antisymmetric response sector. The work
curvature may therefore be viewed as a geometric measure of
the competition between coherent Hamiltonian dynamics and the
basis structure imposed by dissipation. From this perspective,
the emergence of a Poisson sector is not merely a consequence
of nonequilibrium driving; it reflects the inability of the
system to simultaneously diagonalize its energetic and
dissipative structures.

Regions where $F_{\mu\nu}=0$ mark the boundary between
purely metric and Poisson sectors of the response geometry.
In the qubit model studied below, such regions coincide with
parameter regimes where the stationary state becomes aligned
with the Hamiltonian eigenbasis and geometric work
disappears.

\subsection{Toward a K\"{a}hler-Like Response Geometry}
\label{subsec:kahler-response}

The response decomposition developed above reveals that 
stationary-state response possesses two distinct but 
intrinsically linked geometric sectors. Both $G_{\mu\nu}$ 
and $F_{\mu\nu}$ are projections of the same response 
tensor,
\begin{equation}
\chi_{\mu\nu} = G_{\mu\nu} - \frac{1}{2}F_{\mu\nu},
\label{eq:chi_decomp}
\end{equation}
and are therefore not independent structures placed on a 
common manifold but complementary faces of a single 
geometric object. This is the feature that elevates the 
decomposition beyond a bookkeeping exercise.

The coexistence of a symmetric metric sector and a closed 
antisymmetric two-form on the same manifold is the 
defining characteristic of K\"{a}hler geometry. A full 
K\"{a}hler structure requires, in addition, an integrable 
complex structure $J$ satisfying $J^2 = -\mathbf{I}$ that 
is compatible with both $G_{\mu\nu}$ and $F_{\mu\nu}$. We 
do not construct such a $J$ here. The present framework is 
therefore more precisely characterized as an almost 
K\"{a}hler response geometry: the metric and symplectic 
sectors coexist and share a common origin in 
Eq.~(\ref{eq:chi_decomp}), but integrability of the 
complex structure remains an open question.

The physical content of this geometry is sharpest when 
viewed through the partition of the control manifold 
established in Sec.~\ref{sec:response}. When both 
$\lambda^\mu$ and $\lambda^\nu$ are parameters of the 
system Hamiltonian $H$, the antisymmetric sector vanishes 
identically, $F_{\mu\nu} = 0$, and the geometry is purely 
Riemannian. The symplectic sector activates only when the 
control manifold extends into bath parameter space. 
Hamiltonian parameter directions therefore span a purely 
metric submanifold, while bath parameter directions 
introduce the symplectic sector. The control manifold 
carries both structures precisely because open quantum 
systems couple a Hamiltonian sector to a dissipative sector 
that need not share the same preferred basis.

This partition suggests a concrete physical interpretation 
of the missing complex structure. If $J$ exists, it must 
map Hamiltonian parameter directions to bath parameter 
directions, exchanging the purely Riemannian submanifold 
with the symplectic sector. Whether such a $J$ can be 
constructed from the physical data of the problem --- the 
Liouvillian, the stationary state, and the pointer basis 
--- is an open question, but one with a precise geometric 
meaning.

We also note that the natural partition of the control 
manifold into two complementary subspaces is reminiscent 
of para-K\"{a}hler geometry~\cite{Cruceanu:1996}, in which 
the tangent bundle splits into two Lagrangian distributions 
related by a paracomplex structure $K$ satisfying 
$K^2 = +\mathbf{I}$ rather than $J^2 = -\mathbf{I}$. In 
the present setting, the Hamiltonian parameter directions 
and the bath parameter directions play the role of these 
two complementary distributions: the former span the 
purely Riemannian submanifold where $F_{\mu\nu} = 0$, 
while the latter carry the symplectic sector. Whether the 
response geometry of open quantum systems admits a formal 
para-K\"{a}hler structure, and whether the paracomplex 
structure $K$ can be identified with a physical operation 
exchanging system and bath degrees of freedom, are 
questions we leave for future work.

\section{Driven Dissipative Qubit}
\label{sec:model}
To illustrate the response-geometric framework developed in
Sec.~\ref{sec:response}, we consider the minimal open quantum
system capable of exhibiting both reciprocal and
nonreciprocal stationary response: a driven two-level system
subject to pure dephasing. This model was previously used to
investigate geometric work and steady-state curvature in open
quantum systems~\cite{Bittner:2026aa,Bittner:2026ab}. Here,
however, our objective is different. Rather than focusing on
the work performed over a control cycle, we examine the
structure of the stationary-state response tensor itself and
its decomposition into symmetric and antisymmetric sectors.

The model is particularly useful because the origin of the
antisymmetric response can be isolated analytically. No
population relaxation, thermal excitation, gain, or
reservoir engineering is required. Pure dephasing alone is
sufficient to generate finite curvature whenever the basis
preferred by the dissipative dynamics differs from the
Hamiltonian eigenbasis. The resulting stationary state
provides a simple setting in which to examine the emergence
of response geometry, nonreciprocity, and geometric work from
a common underlying mechanism.

As shown in Sec.~\ref{sec:response}, the response tensor
naturally decomposes into a symmetric sector
$G_{\mu\nu}$ and an antisymmetric sector
$F_{\mu\nu}=\chi_{\nu\mu}-\chi_{\mu\nu}$. The present model
provides an explicit realization of this decomposition. We
first construct the stationary state and evaluate the
response tensor directly. We then compute the corresponding
response metric $G_{\mu\nu}$, work curvature
$F_{\mu\nu}$, and Bures information metric
$g_{\mu\nu}^{\mathrm{Bures}}$. Comparison of these quantities
reveals that response geometry and information geometry
characterize distinct aspects of the same nonequilibrium
steady-state manifold.

\subsection{Model and Stationary State}
To illustrate this, we consider a two-level system described by the Hamiltonian
\begin{equation}
H=
\frac{\omega}{2}\sigma_z
+
\frac{g}{2}\sigma_x,
\label{eq:ham}
\end{equation}
where $\omega$ sets the longitudinal energy splitting and
$g$ introduces a transverse field. The Hamiltonian therefore
defines an energy eigenbasis associated with the effective
field
\begin{equation}
\mathbf{h}=(g,0,\omega).
\end{equation}

The system is coupled to a Markovian environment through a
pure-dephasing Lindblad operator
\begin{equation}
L=\sqrt{\gamma}\,\sigma_z,
\label{eq:lindblad_op}
\end{equation}
with dephasing rate $\gamma$. Unlike the Hamiltonian, the
dissipator does not determine energies; instead, it selects
a preferred or pointer basis, namely the eigenstates of
$\sigma_z$, by suppressing coherence between them. The
corresponding master equation is
\begin{equation}
\dot{\rho}
=
-\frac{i}{\hbar}[H,\rho]
+
\gamma
\left(
\sigma_z\rho\sigma_z-\rho
\right).
\label{eq:master}
\end{equation}

The essential feature of this model is the competition 
between the energy basis defined by $H$ and the pointer 
basis selected by the environment. The Hamiltonian 
eigenstates are superpositions of $\sigma_z$ eigenstates 
whenever $g \neq 0$; the environment, however, 
continuously damps coherences in the $\sigma_z$ basis 
regardless of the Hamiltonian structure. The stationary 
state is therefore set by a competition: coherent 
dynamics drive the system toward the energy eigenbasis, 
while dissipation pulls it toward the pointer basis. 
Neither wins outright, and the result is a stationary 
state that retains finite coherence in the energy 
representation.

The degree of misalignment between the two bases is 
characterized by the angle
\begin{equation}
\theta = \tan^{-1}\!\left(\frac{g}{\omega}\right),
\end{equation}
which measures how far the Hamiltonian eigenbasis is 
rotated from the pointer basis in the Bloch sphere. 
At $\theta = 0$ ($g = 0$) the two bases coincide, 
the stationary state is diagonal in the energy 
eigenbasis, and the response is purely reciprocal. 
As $\theta$ increases, basis incompatibility grows, 
stationary-state coherence develops, and --- as we 
show below --- the antisymmetric response sector 
activates. The angle $\theta$ therefore serves as 
a single geometric parameter controlling the 
departure of the response geometry from its purely 
Riemannian, equilibrium-like limit.

The stationary-state response tensor $\chi_{\mu\nu}$ 
can be evaluated directly from the Bloch vector 
components given below in Eqs.~\ref{eq:ss}. 
Its decomposition into symmetric and antisymmetric 
sectors then provides a concrete realization of the 
framework developed in Sec.~\ref{sec:response}: the 
symmetric sector $G_{\mu\nu}$ characterizes reciprocal 
susceptibility, while the antisymmetric sector 
$F_{\mu\nu}$ measures the failure of reciprocity and 
carries the geometric work. The central question is 
whether basis incompatibility --- parametrized by 
$\theta$ --- is sufficient on its own to generate 
a finite antisymmetric sector, without population 
relaxation, gain, or reservoir engineering.

Writing the stationary state in Bloch form,
\begin{equation}
\rho_{\mathrm{ss}}
=
\frac12
\left(
I+\mathbf r_{\mathrm{ss}}\cdot\boldsymbol{\sigma}
\right),
\label{eq:ss_density}
\end{equation}
with
\begin{equation}
\mathbf r_{\mathrm{ss}}
=
(x_{\mathrm{ss}},
y_{\mathrm{ss}},
z_{\mathrm{ss}}),
\end{equation}
the steady-state solution of
Eq.~(\ref{eq:master}) is
\begin{align}
x_{\mathrm{ss}}
&=
\frac{2s\,\omega g}
{\gamma D},
\\
y_{\mathrm{ss}}
&=
-\frac{s\,g}
{D},
\\
z_{\mathrm{ss}}
&=
\frac{s\,(4\omega^{2}+\gamma^{2})}
{2\gamma D},
\label{eq:ss}
\end{align}
where
\begin{equation}
D
=
2\omega^{2}
+
g^{2}
+
\frac{\gamma^{2}}{2},
\label{eq:d}
\end{equation}
and $s$ denotes the stationary polarization parameter.
Several features are immediately apparent. When $g=0$, the
off-diagonal Bloch components vanish and the stationary
state is aligned with the Hamiltonian eigenbasis. 

For
$g\neq0$, the components $x_{\mathrm{ss}}$ and
$y_{\mathrm{ss}}$ become nonzero, signaling the appearance
of stationary-state coherence. As we show below, the same
parameter that generates coherence also activates the
antisymmetric response sector and the associated work
curvature. In this sense, the qubit model makes explicit
the connection between basis incompatibility, coherence,
and geometric response.

\subsection{Response Tensor and Geometric Decomposition}
The generalized forces conjugate to the control parameters
$\omega$ and $g$ are obtained from the derivatives of the
Hamiltonian,
\begin{equation}
O_\omega
=
\frac{\partial H}{\partial \omega}
=
\frac{\sigma_z}{2},
\qquad
O_g
=
\frac{\partial H}{\partial g}
=
\frac{\sigma_x}{2}.
\end{equation}
The corresponding components of the work connection are
\begin{equation}
A_\omega
=
\Tr(\rho_{\mathrm{ss}} O_\omega)
=
\frac{z_{\mathrm{ss}}}{2},
\qquad
A_g
=
\Tr(\rho_{\mathrm{ss}} O_g)
=
\frac{x_{\mathrm{ss}}}{2}.
\label{eq:A_components}
\end{equation}
The stationary-state response tensor is therefore
\begin{equation}
\chi_{\mu\nu}
=
\partial_\nu A_\mu,
\label{eq:chi_def_model}
\end{equation}
or explicitly,
\begin{equation}
\chi
=
\begin{pmatrix}
\partial_\omega A_\omega &
\partial_g A_\omega
\\[6pt]
\partial_\omega A_g &
\partial_g A_g
\end{pmatrix}.
\label{eq:chi_matrix}
\end{equation}

Following Sec.~\ref{sec:response}, we decompose the response
tensor into symmetric and antisymmetric sectors,
\begin{equation}
G_{\mu\nu}
=
\frac12
\left(
\chi_{\mu\nu}
+
\chi_{\nu\mu}
\right),
\label{eq:G_model}
\end{equation}
and
\begin{equation}
F_{\mu\nu}
=
\chi_{\nu\mu}
-
\chi_{\mu\nu}.
\label{eq:F_model}
\end{equation}
For the present two-parameter model, the only independent
antisymmetric component is
\begin{equation}
F_{\omega g}
=
\chi_{g\omega}
-
\chi_{\omega g}
=
\partial_\omega A_g
-
\partial_g A_\omega .
\end{equation}
Substituting Eq.~(\ref{eq:ss}) into
Eq.~(\ref{eq:A_components}) yields
\begin{equation}
F_{\omega g}
=
\frac{s\,g\,(g^2+\gamma^2)}
{\gamma D^2},
\label{eq:F_closed}
\end{equation}
where $D$ is given by Eq.~(\ref{eq:d}).

Several important features follow immediately from
Eq.~(\ref{eq:F_closed}). First, the curvature vanishes when
$g=0$, corresponding to perfect alignment between the
Hamiltonian and pointer bases. In this limit the stationary
state is diagonal in the energy basis and the response is
locally integrable. Second, the curvature is odd in $g$,
indicating that the orientation of response circulation
changes upon reversing the direction of the transverse field.
Finally, the magnitude of the curvature grows with increasing
basis misalignment, demonstrating that stationary-state
coherence and antisymmetric response arise from the same
underlying mechanism.

\begin{figure*}[t]
\centering
\includegraphics[width=\textwidth]{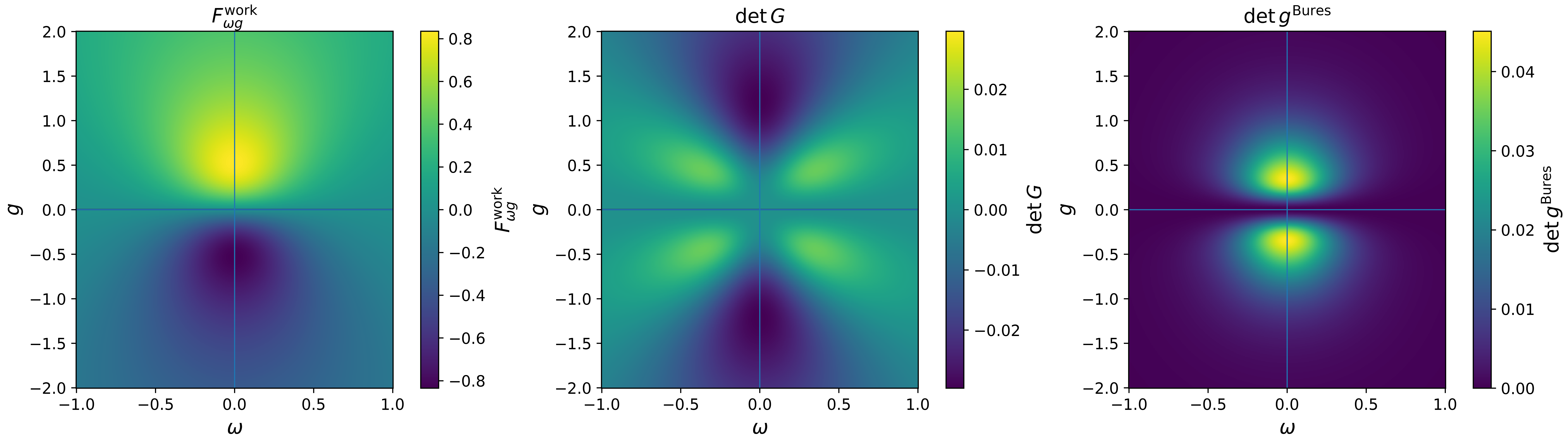}
\caption{
\label{fig:1}
Response geometry of the driven dissipative qubit in the
control-parameter plane $(\omega,g)$ for fixed dissipation
rate $\gamma=1$ and stationary polarization parameter
$s=0.75$. Left: antisymmetric response tensor (work
curvature) $F_{\omega g}=\chi_{g\omega}-\chi_{\omega g}$.
The curvature changes sign across the symmetry line $g=0$
and vanishes when the stationary state becomes aligned with
the Hamiltonian eigenbasis. Center: determinant of the
symmetric response tensor
$G_{\mu\nu}=\tfrac12(\chi_{\mu\nu}+\chi_{\nu\mu})$,
characterizing the reciprocal susceptibility structure of
the steady-state manifold. Right: determinant of the Bures
information metric $g_{\mu\nu}^{\rm Bures}$, which measures
the statistical distinguishability of neighboring steady
states. The distinct spatial organization of the response
and information metrics illustrates that reciprocal response
and state distinguishability capture different geometric
aspects of the same nonequilibrium steady-state manifold.
}
\end{figure*}

Figure~\ref{fig:1} presents the resulting response geometry
over the control-parameter plane $(\omega,g)$. The left
panel shows the antisymmetric response tensor
$F_{\omega g}$, the center panel shows the determinant of
the symmetric response tensor $G_{\omega g}$, and the right
panel shows the determinant of the Bures information metric
for comparison.
Several observations emerge immediately. The antisymmetric
response vanishes identically along the line $g=0$, where
the Hamiltonian and pointer bases coincide. In this limit
the stationary state is diagonal in the energy basis and
the response becomes locally integrable. Away from this
line, finite curvature develops with opposite signs for
positive and negative values of $g$. The sign change
reflects a reversal in the orientation of response
circulation on the control manifold and is a direct
manifestation of nonreciprocity,
$\chi_{\omega g}\neq\chi_{g\omega}$.

The central panel displays the determinant of the symmetric
response tensor,
\[
\det G
=
G_{\omega\omega}G_{gg}
-
G_{\omega g}^2.
\]
Unlike the curvature $F_{\omega g}$, which changes sign
under $g\rightarrow -g$, the symmetric response is governed
by the reciprocal susceptibilities
$G_{\omega\omega}$, $G_{gg}$, and the mixed response
$G_{\omega g}=G_{g\omega}$. These quantities characterize
how the steady state responds to parameter variations
without regard to orientation and therefore contain no
information about response circulation or geometric work.
The tensors $G_{\omega g}$ and $F_{\omega g}$ encode
complementary aspects of stationary-state response.
The symmetric component $G_{\omega g}$ measures reciprocal
coupling between the control directions $\omega$ and $g$,
whereas the antisymmetric component $F_{\omega g}$
quantifies the failure of reciprocity and determines the
local work curvature.

Figure~\ref{fig:1} makes the decomposition 
concrete. The left panel shows that the antisymmetric 
response $F_{\omega g}$ is not a global property of 
the manifold but a local one: it develops where the 
Hamiltonian and pointer bases are misaligned and 
vanishes identically along $g = 0$ where they coincide. 
The center panel shows that the symmetric sector 
$G_{\mu\nu}$ has no such restriction --- it remains 
finite across the manifold, reflecting that reciprocal 
susceptibility persists regardless of basis alignment. 
The two sectors therefore partition the response into 
physically distinct contributions: $G_{\mu\nu}$ 
measures how the steady state responds to parameter 
variations, while $F_{\mu\nu}$ measures whether that 
response is reciprocal. Crucially, both arise from 
the same tensor $\chi_{\mu\nu}$ and require no 
separate geometric construction. The origin of the 
curvature is not exotic: a single Hamiltonian parameter 
$g$ tilting the energy eigenbasis away from the 
pointer basis is sufficient. No population relaxation, 
pumping, or reservoir engineering is needed. Geometric 
work emerges wherever two competing structures --- 
the Hamiltonian eigenbasis and the environment-selected 
pointer basis --- fail to organize the steady-state 
manifold in the same way.

\subsection{Response Geometry versus Information Geometry}


The response tensor $\chi_{\mu\nu}$ characterizes what 
happens when the system is perturbed: how generalized 
forces shift in response to parameter variations. A 
conceptually distinct question is how rapidly neighboring 
steady states become experimentally distinguishable --- 
that is, how much the stationary state itself changes as 
parameters vary. This second question is answered by the 
Bures metric $g^{\rm Bures}_{\mu\nu}$, which measures 
the statistical distance between neighboring density 
matrices~\cite{Bures1969,Uhlmann1976,BraunsteinCaves1994,
ProvostVallee1980,Paris2009} and sets the fundamental 
limit on parameter estimation via quantum Fisher 
information.

These two geometric structures need not agree, and in 
general they do not. A system can be highly sensitive 
to parameter variations (large $G_{\mu\nu}$) while 
neighboring steady states remain difficult to distinguish 
(small $g^{\rm Bures}_{\mu\nu}$), or vice versa. The 
right panel of Fig.~\ref{fig:1} makes this 
explicit: the spatial organization of 
$\det g^{\rm Bures}$ is markedly different from that 
of $\det G$, even though both are computed from the 
same family of stationary states. Response geometry 
characterizes what the system \emph{does} when 
perturbed; information geometry characterizes how 
readily neighboring states can be told apart. Both 
are physically meaningful, but they are not 
interchangeable.

\subsection{Basis Incompatibility and the Origin of 
Geometric Response}

The qubit model isolates the minimal ingredients for 
geometric response: a Hamiltonian and a dissipator that 
select incompatible preferred bases. No population 
relaxation, exceptional points, or reservoir engineering 
is required. The angle $\theta$ alone controls the 
departure from reciprocal response, and $F_{\mu\nu}$ 
vanishes precisely when $\theta = 0$.

This simplicity points toward a general principle. 
Geometric response emerges whenever the energetic and 
dissipative structures of an open quantum system fail 
to share a common eigenbasis. In molecular and 
condensed-phase systems, this situation is the rule 
rather than the exception. Electronic excited states 
are organized by the molecular Hamiltonian, while 
vibrational relaxation and pure dephasing are governed 
by the nuclear environment --- two structures that 
generically point in different directions in Liouville 
space. In excitonic systems, the basis diagonalizing 
the Frenkel Hamiltonian need not coincide with the 
basis selected by site-local fluctuations. In 
cavity-polariton systems, light-matter coupling mixes 
the photonic and excitonic components of the polariton 
basis while the reservoir acts on each component 
differently.

In each case the antisymmetric response sector 
activates and geometric work becomes nonzero. The 
curvature $F_{\mu\nu} = \chi_{\nu\mu} - \chi_{\mu\nu}$ 
is directly accessible from cross-response measurements 
--- no state tomography required. The qubit result 
therefore does more than illustrate the framework; it 
identifies a mechanism broadly operative in open quantum 
systems of chemical and physical interest.

\section{Discussion}
\label{sec:discussion}

\subsection{Response Geometry Beyond Equilibrium 
Thermodynamics}

The identification $F_{\mu\nu} = \chi_{\nu\mu} - 
\chi_{\mu\nu}$ has a consequence that extends beyond 
the specific model analyzed here. It establishes that 
the geometry of thermodynamic response is not fixed 
by equilibrium structure but is determined by the 
symmetry properties of the response tensor itself. 
In equilibrium, reciprocity is exact and the geometry 
is purely Riemannian --- the Weinhold and Ruppeiner 
constructions are the complete story. Open quantum 
systems break this restriction not by adding new 
geometric structure from outside, but by activating 
a sector of response geometry that reciprocity had 
suppressed.

This reframing has practical implications. The work 
curvature is not a separate object to be computed 
alongside the response tensor --- it is already 
contained within it. Measuring $\chi_{\mu\nu}$ and 
$\chi_{\nu\mu}$ independently is sufficient to 
extract $F_{\mu\nu}$ without any additional 
geometric construction. Conversely, a system that 
appears reciprocal --- $\chi_{\mu\nu} = \chi_{\nu\mu}$ 
--- has vanishing geometric work by necessity, 
regardless of how complex its dynamics may be. 
Reciprocity is therefore not merely a symmetry 
property of the response tensor; it is the 
geometric condition that determines whether 
thermodynamic work can accumulate over a closed 
control cycle.

\subsection{Basis Incompatibility and the Origin of 
Geometric Response}

The qubit model identifies basis incompatibility as 
the essential ingredient for geometric response. This 
is a stronger statement than it might appear. Many 
open quantum systems retain stationary-state coherence 
while remaining effectively reciprocal --- coherence 
alone does not activate the antisymmetric sector. The 
crucial combination is coherence \emph{and} basis 
incompatibility: the stationary state must be unable 
to simultaneously satisfy the organizations imposed 
by the Hamiltonian and the environment. When that 
condition holds, reciprocity breaks and $F_{\mu\nu}$ 
becomes finite.

This condition is broadly satisfied in molecular and 
condensed-phase systems. Competing bases arise 
naturally wherever optical excitation, energy 
transport, and environmental relaxation act on 
different preferred representations --- bright and 
dark excitonic states, localized and delocalized 
representations, Hamiltonian eigenstates versus 
site-local fluctuation bases. In each case the 
response curvature $F_{\mu\nu}$ provides a 
quantitative measure of the extent to which these 
competing organizations fail to admit a common 
description, and is directly extractable from 
cross-response measurements.

\subsection{Response Geometry and Information Geometry}

The distinction between response geometry and 
information geometry is not merely technical. 
Information geometry --- embodied by the Bures 
metric --- answers the question of how rapidly 
neighboring steady states become distinguishable 
as parameters vary. Response geometry answers the 
question of how the system reacts when perturbed. 
These are different experiments, and there is no 
general reason they should agree.

The fact that they do not coincide in the qubit 
model has a direct implication: knowing the 
information geometry of a steady-state manifold 
is not sufficient to determine its response 
geometry, and vice versa. A complete geometric 
characterization of a nonequilibrium steady state 
requires both structures. Whether deeper 
compatibility conditions connect them --- analogous 
to the way metric and symplectic structures are 
linked in K\"{a}hler geometry --- remains an open 
question, and one that may be particularly 
tractable in systems where the response tensor 
admits an almost K\"{a}hler decomposition as 
identified in Sec.~\ref{subsec:kahler-response}.

\subsection{Experimental Accessibility and Outlook}

The work curvature is directly observable. Since 
$F_{\mu\nu} = \chi_{\nu\mu} - \chi_{\mu\nu}$, it 
is extracted from cross-response measurements 
without state tomography or spectral decomposition 
of the Liouvillian. Potential platforms include 
driven spin systems probed by ESR or NMR, cavity 
and circuit QED architectures, and excitonic and 
polaritonic materials --- any driven quantum system 
in which coherent dynamics compete with dissipative 
processes and optical excitation, transport, and 
relaxation define distinct preferred bases.

A useful physical analogy is provided by anisotropic
optical media. Ordinary birefringence arises from a
symmetric dielectric tensor and therefore remains
reciprocal, producing distinct propagation velocities
for different polarizations without generating an
antisymmetric response sector. Magneto-optic media,
by contrast, acquire an antisymmetric contribution
to the dielectric response that produces Faraday
rotation and nonreciprocal polarization transport.
Ordinary birefringence is therefore analogous to the
metric sector $G_{\mu\nu}$, whereas gyrotropic and
magneto-optic effects are the natural optical
analogue of $F_{\mu\nu}$. Both cases illustrate how
competing structures acting on the same manifold
generate geometric transport phenomena --- the
difference is that in open quantum systems the
competing structures are the Hamiltonian eigenbasis
and the environment-selected pointer basis rather
than crystal symmetry and an applied magnetic field.

Several open questions emerge from this work. How 
does the response geometry develop in interacting 
many-body systems where collective correlations 
generate nontrivial steady-state structure? Can 
response curvature serve as a diagnostic of 
nonequilibrium functionality, identifying regimes 
where competing energetic and dissipative structures 
enhance transport, sensing, or energy conversion? 
And does the almost K\"{a}hler structure identified 
in Sec.~\ref{subsec:kahler-response} admit a formal completion 
--- a complex structure $J$ mapping Hamiltonian 
parameter directions to bath parameter directions 
--- that would place the response geometry of open 
quantum systems within a fully K\"{a}hler framework?

Beyond quasistatic thermodynamics, the 
response-geometric framework connects naturally to 
spectroscopic response theory. In nonlinear 
spectroscopy, the measured signal is a sum of 
Liouville pathway amplitudes propagated between 
successive field interactions. These pathways are 
not static: the system evolves continuously under 
coherent dynamics and environmental fluctuations 
between interactions, continuously transporting 
amplitude through Liouville space. The antisymmetric 
sector of the response tensor measures the extent 
to which distinct pathway transport mechanisms fail 
to commute --- a noncommuting pathway transport that 
leaves direct signatures in multidimensional spectra. 
In an accompanying paper we develop this perspective 
explicitly in the context of two-dimensional 
electronic spectroscopy, where pathway transport by 
coherent and dissipative dynamics can be related 
directly to experimentally observable spectral 
features.

\vspace{0.5cm}

\section*{Data availability Statement}
The numerical data and code that support the findings of this study are openly available in the Borealis Dataverse Repository at http://doi.org/[doi], reference number [reference number will be added before publication].

\begin{acknowledgments}
CSA acknowledges funding from the Government of Canada (Canada Excellence Research Chair CERC-2022-00055),  
from the Institut Courtois, Facult\'e des arts et des sciences, Universit\'e de Montr\'eal (Chaire de recherche de direction de l'Institut Courtois) and from the Natural Science and Engineering Research Council of Canada (NSERC Discovery Grant RGPIN-2024-05893). 
ERB acknowledges funding from the National Science Foundation (CHE-2404788), Robert A.\ Welch Foundation (E-1337), the Department of Energy supported this research through Award No. 11937-PO147716. ERB gratefully acknowledges funding from IVADO for a Visiting Professorship at the Institut Courtois, Universit\'e de Montr\'eal. 
\end{acknowledgments}

\subsection*{Use of Generative Artificial Intelligence}
In compliance with institutional guidelines of the Universit\'e de Montr\'eal, generative artificial intelligence tools were used to assist with the editing of language and stylistic refinement of parts of the manuscript and to assist in the synthesis of the literature. These tools were not used to generate scientific content, perform analysis, or influence the interpretation of results. All content has been reviewed and validated by the authors, who assume full responsibility for the manuscript.


\bibliographystyle{apsrev4-2}
\bibliography{refs_consolidated}

\end{document}